\documentclass[preprint]{aastex}
\usepackage{natbib}

\newcommand{\me}{\mathrm{e}}   
\newcommand{\mi}{\mathrm{i}}   
\newcommand{\sinc}{{\rm sinc}} 
\newcommand{\sm}{\ifmmode\mbox{M}_{\odot}\else$\mbox{M}_{\odot}$\fi}

\shortauthors{Ransom, Cordes and Eikenberry}
\shorttitle{A Binary Pulsar Search Technique}

\begin{document}

\title{A New Search Technique for Short Orbital Period Binary Pulsars}

\author{Scott M. Ransom\altaffilmark{1,2}} 

\author{James M. Cordes\altaffilmark{3}}

\and 

\author{Stephen S. Eikenberry\altaffilmark{3}}

\altaffiltext{1}{Department of Physics, McGill University, Montreal,
  QC. H3A~2T8 Canada}
\altaffiltext{2}{Center for Space Research, Massachusetts Institute of
  Technology, Cambridge, MA 02139}
\altaffiltext{3}{Astronomy Department, Cornell University, Ithaca, NY 14853}

\begin{abstract}
  We describe a new and efficient technique which we call sideband or
  phase-modulation searching that allows one to detect short period
  binary pulsars in observations longer than the orbital period.  The
  orbital motion of the pulsar during long observations effectively
  modulates the phase of the pulsar signal causing sidebands to appear
  around the pulsar spin frequency and its harmonics in the Fourier
  transform.  For the majority of binary radio pulsars or Low-Mass
  X-ray Binaries (LMXBs), large numbers of sidebands are present,
  allowing efficient searches using Fourier transforms of short
  portions of the original power spectrum.  Analysis of the complex
  amplitudes and phases of the sidebands can provide enough
  information to solve for the Keplerian orbital parameters.  This
  technique is particularly applicable to radio pulsar searches in
  globular clusters and searches for coherent X-ray pulsations from
  LMXBs and is complementary to more standard ``acceleration''
  searches.
\end{abstract}


\section{Introduction}

Short orbital-period binary pulsars have proven to be sensitive
laboratories that allow tests of a variety of physical processes
including theories of gravitation.  Unfortunately, except for those
systems containing the brightest pulsars, they are notoriously
difficult to detect.  When the observation time $T_{obs}$ is greater
than a small fraction of the orbital period $P_{orb}$, Doppler effects
due to the orbital motion of the pulsar cause drastic reductions in
the sensitivities of pulsar searches \citep{jk91}.

Numerous groups \citep[e.g.][]{mk84, agk+90, whn+91} have developed
variations of so called ``acceleration'' searches in order to mitigate
the loss in sensitivity caused by orbital motion.  These searches can
almost completely recover this lost sensitivity when $T_{obs} \lesssim
P_{orb}/10$ by taking advantage of the fact that orbital motion causes
an approximately linear drift of the pulsar spin frequency $f_{spin}$.
Recent acceleration searches of globular clusters have discovered
numerous binary systems, some with orbital periods as short as
$\sim90$\,min \citep{clf+00, rgh+01, dlm+01}.

Acceleration searches are effective only for $T_{obs} \lesssim
P_{orb}/10$, but search sensitivities improve as $T_{obs}^{1/2}$.  In
effect, acceleration searches are limited to discovering only the
brightest binary pulsars in ultra-compact --- systems with orbital
periods less than a few hours --- orbits.  For a 2\,ms pulsar in a
1\,hour orbit with a low mass white dwarf (WD) companion, the optimal
observation time $T_{best}$ for an acceleration search \citep[using
eqn.~22 from][]{jk91} is only $\sim300-600$ seconds.  If this pulsar
were located in a globular cluster, where observations of
$5-10$\,hours in duration are common, acceleration searches would be a
factor of $\sqrt{T_{obs}/T_{best}}\sim8$ times less sensitive than an
optimal (i.e.~coherent) search of a full observation.

We have developed an efficient search technique for short orbital
period binary pulsars that complements the use of acceleration
searches.  Since it requires $T_{obs} \gtrsim 1.5P_{orb}$ and its
sensitivity improves as the ratio $r_{orb} = T_{obs} / P_{orb}$
\emph{increases}, it allows the search of very interesting portions of
orbital parameter space for the first time.  This technique was
developed and tested over the past several years as part of a
Ph.D.~thesis \citep{ran01}.  A similar but independent derivation of
many of the properties of this technique has been recently published
by \citet{jrs+02}, based on an earlier mention of this work in
\citet{ran00}.  While the basic search techniques presented by these
papers are nearly identical, aspects concerning the determination of
the orbital elements of a candidate pulsar discovered with this
technique are presented in more detail in this work.


\section{Phase Modulation due to Orbital Motion} \label{sec:phsmod}

During observations of a binary pulsar where $T_{obs}$ is longer than
$P_{orb}$, the orbital motion causes a modulation of the phase of the
pulsar signal.  Differences in light travel time across the projected
orbit advance or delay the pulse phase in a periodic fashion.

\subsection{Circular Orbits} \label{sec:circular}

If the orbit is circular (where the eccentricity $e$ and the angle of
periapsis $\omega$ are both zero), the phase delays are sinusoidal.
The fundamental harmonic\footnote{The following analysis can be
  applied to any harmonic $h$ of a pulsar's signal by substituting
  $hr_{spin}$ for $r_{spin}$ and the appropriate harmonic phase for
  $\phi_{spin}$.} of the resulting signal as sampled at a telescope
can be described by
\begin{equation}
  \label{eq:pmsig}
  n_j = n(u) = a \,\cos\left[2\pi r_{spin}u + \phi_{spin} +
  \Phi_{orb}\cos(2\pi r_{orb}u + \phi_{orb})\right],
\end{equation}
where $a$ and $\phi_{spin}$ are the amplitude and phase of the
pulsation, $\Phi_{orb}$ and $\phi_{orb}$ are the amplitude and phase
of the modulation due to the orbit, $r$ is a Fourier frequency
(i.e.~$r = fT_{obs} = T_{obs}/P$), and $u = j/N = (t_j-t_o)/T_{obs}$
is the discrete dimensionless time during the observation of the
$j^{th}$ sample of $N$ such that $0 \le u \le 1$.  We can write
$\Phi_{orb}$ and $\phi_{orb}$ in terms of the three Keplerian orbital
parameters for circular orbits ($P_{orb}$, the semi-major axis
$x_{orb}=a_1\sin(i)/c$, and the time of periapsis $T_o$) as
\begin{equation}
  \label{eq:Phiorb}
  \Phi_{orb} = 2\pi x_{orb}f_{spin} \simeq 
  7.375\,\frac{f_1^{1/3}P_{orb}^{2/3}({\rm hr})}{P_{spin}({\rm sec})}
\end{equation}
and
\begin{equation}
  \label{eq:phiorb}
  \phi_{orb} = 2\pi\frac{t_o-T_o}{P_{orb}} + \frac{\pi}{2},
\end{equation}
where the units of both are radians, and the pulsar's mass function
$f_1$ is defined as
\begin{equation}
  \label{eq:massfunct}
  f_1=\frac{\left[M_2\sin(i)\right]^3}{\left(M_1+M_2\right)^2},
\end{equation}
where $i$ is the orbital inclination and $M_1$ and $M_2$ are the
pulsar and companion masses (in solar masses) respectfully.

The Discrete Fourier Transform (DFT) of the signal in
eqn.~\ref{eq:pmsig} when $\Phi_{orb}=0$ can be approximated as
\citep*[see e.g.~][]{rem02}
\begin{mathletters}
  \begin{eqnarray}
    A(r) 
    \label{eq:ftdef}
    & = & \sum_{j=0}^{N-1} n_j \,\me^{-2\pi\mi jr/N}
    \simeq N\int_{0}^{1} n(u) \,\me^{-2\pi\mi ru}\,\mathrm{d}u \\
    \label{eq:cosint}
    & \simeq & \frac{aN}{2}\,\me^{\mi\phi_{spin}}
    \int^1_0 \me^{2\pi\mi (r_{spin}-r) u}\,\mathrm{d}u \\
    & = & A_{coherent}\,\me^{\mi\pi (r_{spin}-r)}
    \frac{\sin\left[\pi(r_{spin}-r)\right]}{\pi(r_{spin}-r)},
   \end{eqnarray}
\end{mathletters}%
where $A_{coherent} = \frac{aN}{2}\,\me^{\mi\phi_{spin}}$ represents the
coherent Fourier response at the spin frequency $r_{spin}$.
Similarly, we can approximate the DFT of our phase modulated signal
from eqn.~\ref{eq:pmsig} as
\begin{mathletters}
  \begin{eqnarray}
    A(r)
    & = & \frac{aN}{2} \int^1_0 \me^{\mi\left[2\pi r_{spin}u + \phi_{spin} + 
      \Phi_{orb}\cos(2\pi r_{orb}u + \phi_{orb})\right]}\,
  \me^{-2\pi \mi ru}\,\mathrm{d}u \\
    \label{eq:pmintegral}
    & = & A_{coherent}\int^1_0 \me^{\mi\Phi_{orb}\cos(2\pi 
      r_{orb}u+\phi_{orb})}\,\me^{2\pi\mi(r_{spin}-r)u}\,\mathrm{d}u.
  \end{eqnarray}
\end{mathletters}%

Using the Jacobi-Anger expansion \citep[e.g.][]{arf85}
\begin{equation}
  \label{eq:Jexpand}
  \me^{\mi z \cos(\theta)} = 
  \sum_{m=-\infty}^{\infty}\mi^mJ_m(z)\,\me^{\mi m\theta} = 
  \sum_{m=-\infty}^{\infty}J_m(z)\,
  \me^{\mi m \left(\theta+\frac{\pi}{2}\right)},
\end{equation}
where $J_m(z)$ is an integer order Bessel function of the first kind,
we can expand the integrand in eqn.~\ref{eq:pmintegral} to give
\begin{mathletters}
  \begin{eqnarray}
    A(r) 
    & = & A_{coherent}\sum_{m=-\infty}^{\infty}J_m(\Phi_{orb})
    \int^1_0 \me^{\mi m(2\pi r_{orb}u + \phi_{orb} + \frac{\pi}{2})}\,
    \me^{2\pi\mi(r_{spin}-r)u}\,\mathrm{d}u \\
    \label{eq:pmfullint}
    & = & A_{coherent}\sum_{m=-\infty}^{\infty}J_m(\Phi_{orb})\,
    \me^{\mi m\left(\phi_{orb}+\frac{\pi}{2}\right)}\int^1_0
    \me^{2\pi\mi\left(r_{spin}+mr_{orb}-r\right)u}\,\mathrm{d}u.
  \end{eqnarray}
\end{mathletters}%
Since the integrand in eqn.~\ref{eq:pmfullint} is identical in form to
that in eqn.~\ref{eq:cosint}, the Fourier transform of a phase
modulated signal is equivalent to the Fourier transform of a series of
cosinusoids at frequencies centered on the pulsar spin frequency
$r_{spin}$, but separated from $r_{spin}$ by $mr_{orb}$ Fourier bins
(i.e.~sidebands).  The response is therefore
\begin{equation}
  \label{eq:pmresp}
  A(r) = A_{coherent}\sum_{m=-\infty}^{\infty}
  J_m(\Phi_{orb})\,\me^{\mi m\left(\phi_{orb}+\frac{\pi}{2}\right)}\,
  \me^{\mi\pi (r_{spin}+mr_{orb}-r)}\,
    \frac{\sin\left[\pi(r_{spin}+mr_{orb}-r)\right]}{\pi(r_{spin}+mr_{orb}-r)}.
\end{equation}

\begin{figure}
  \plottwo{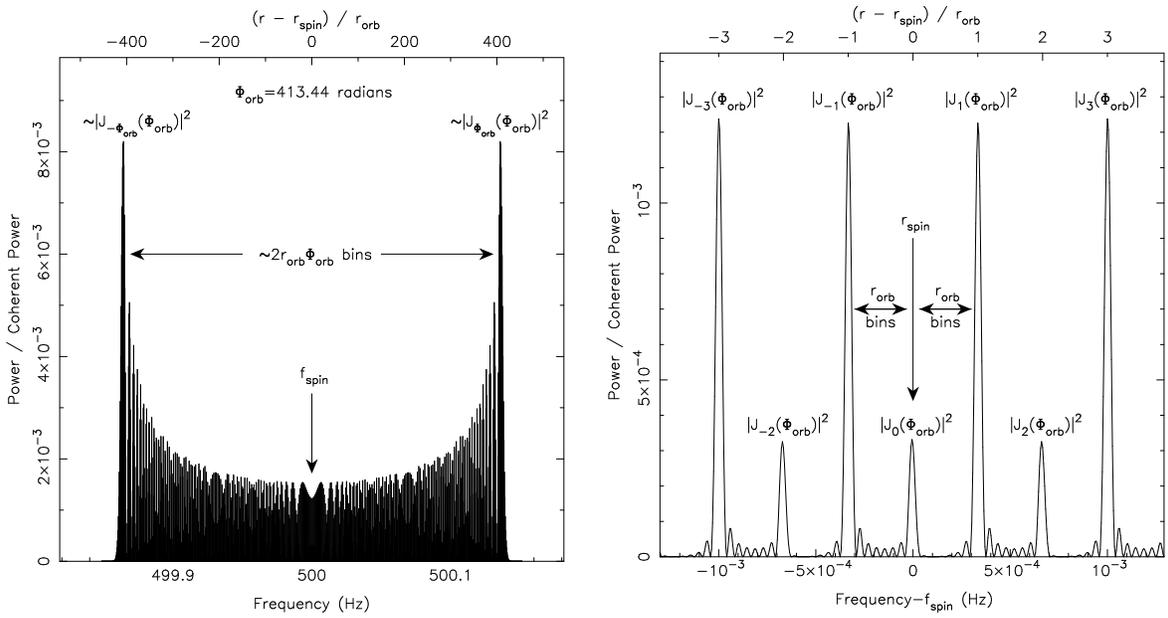}{zoom_phasemod.eps}
  \caption{\footnotesize
    These two panels show the Fourier response of the fundamental spin
    harmonic of a phase modulated 2\,ms binary pulsar.  The properties
    of the response are typical for circular orbits with $\Phi_{orb} =
    2\pi x_{orb}f_{spin} \gg 1$.  The response was calculated for an
    8\,hour observation of a NS-WD binary with a 50\,min orbital
    period, projected semi-major axis $x_{orb} = 0.132$\,lt-s, orbital
    inclination $i=60$\degr, and a companion mass of $\sim0.2\,\sm$.
    The panel on the right shows the central portion of the full
    response on the left.  The periodic nature of the sidebands with
    peak-to-peak spacing (in Fourier bins) of $r_{orb} = T_{obs} /
    P_{orb}$ is obvious.
    \label{fig:phsmod}}
\end{figure}

When $r_{orb}$ is an integer (i.e.~the observation covers an integer
number of complete orbits), the Fourier response shown in
eqn.~\ref{eq:pmresp} becomes particularly simple at Fourier
frequencies $r_{spin}+sr_{orb}$, where $s$ is an integer describing
the sideband in question.  In this case, the summation collapses to a
single term when $m=s$, giving
\begin{equation}
  \label{eq:pmsimple}
  A\left(r_{spin}+sr_{orb}\right)
  = A_{coherent}J_s(\Phi_{orb})\,\me^{\mi s
    \left(\phi_{orb}+\frac{\pi}{2}\right)}.
\end{equation}
In effect, the phase modulated signal produces sidebands composed of a
series of sidebands split in frequency from $r_{spin}$ by $sr_{orb}$
Fourier bins, with amplitudes proportional to $J_s(\Phi_{orb})$, and
phases of $s\left(\phi_{orb}+
  \frac{\pi}{2}\right)+\phi_{spin}$\,radians\footnote{Note that
  $J_{-n}(z) = (-1)^nJ_n(z)$, and also that $J_n(z)$ can be negative.
  If the sideband amplitude determined by the Bessel function is
  negative (i.e.~$J_s(z)$ is negative), the \emph{measured}\ phase of
  the Fourier response will differ by $\pi$\,radians from the
  predicted phase since the \emph{measured}\ amplitude is always
  positive.}.  Figure~\ref{fig:phsmod} shows the Fourier response for
a ``typical'' pulsar-WD system in a circular 50\,min orbit during an
8\,hour observation.

\placefigure{fig:phsmod}

\begin{figure}
  \plotone{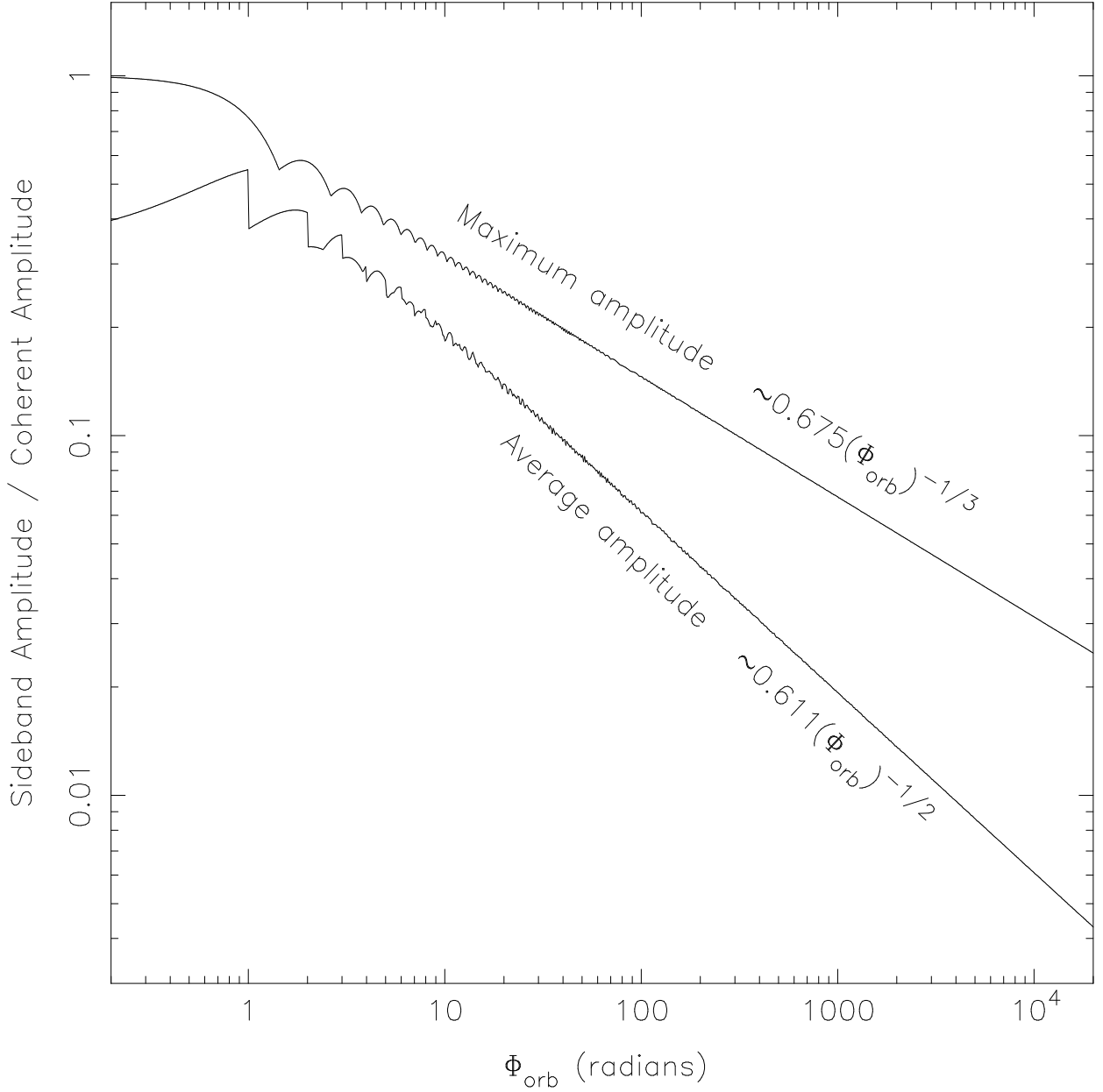}
  \caption{\footnotesize
    Results of the measurement of the maximum and average values of
    $|J_s(\Phi_{orb})|$ where $s$ are all the integers from $0 \to
    \lceil\Phi_{orb}\rceil$.  The Bessel functions correspond to the
    magnitude of the $s^{\rm th}$ sideband pair of the Fourier
    response of a phase modulated sinusoid as a fraction of the
    unmodulated or fully coherent amplitude (see
    \S\ref{sec:circular}).  Approximate asymptotic relations for large
    values of $\Phi_{orb}$ are given above the measured values.
    \label{fig:besj}}
\end{figure}

While the number of sidebands implied by eqn.~\ref{eq:pmresp} is
infinite (assuming $\Phi_{orb} \ne 0$), properties of the Bessel
functions --- which define the magnitude of the Fourier amplitudes of
the sidebands --- give a rather sharp cutoff at $\sim2\Phi_{orb}$
sidebands (or $\sim\Phi_{orb}$ pairs of sidebands).  The maximum value
of $J_s(\Phi_{orb})$ also occurs near $s = \Phi_{orb}$ at a value of
$\sim0.675\Phi_{orb}^{-1/3}$, which gives the phase modulation
response its distinctive ``horned'' shape when $\Phi_{orb}$ is large
(see Fig.~\ref{fig:phsmod}).  The average magnitude of the phase
modulation sidebands falls off more quickly as
$\sim0.611\Phi_{orb}^{-1/2}$ (see Fig.~\ref{fig:besj}).

For pulsars with narrow pulse shapes and many Fourier harmonics, phase
modulation produces sidebands split by $r_{orb}$ Fourier bins centered
around each spin harmonic.  For spin harmonic number $h$, the phase
modulation amplitude is $\Phi_{orb,h} = 2\pi x_{orb}hf_{spin}$.
Therefore, for large $\Phi_{orb,h}$, there are $\sim h$ times more
sidebands than around the fundamental, but the average amplitudes are
smaller by a factor of $\sim h^{-1/2}$.

\placefigure{fig:besj}

\subsection{Small Amplitude Limit} \label{sec:smallamp}

When $\Phi_{orb} \ll 1$, a phase modulated signal will have only three
significant peaks in its Fourier response.  These peaks are located at
$r_{spin}$ and $r_{spin} \pm r_{orb}$ and have magnitudes proportional
to $J_0(\Phi_{orb}) \sim 1 - \Phi_{orb}^2/4 \sim 1$ and
$J_{\pm1}(\Phi_{orb}) \sim \pm \Phi_{orb}/2$ respectively.

\citet{mmn+81} performed a detailed analysis of phase modulated
optical pulsations in the small amplitude limit from the 7.7\,second
X-ray pulsar 4U~1626-67 in order to solve for the orbital parameters
of the system.  The separation of the sidebands provided a measurement
of the $\sim42$\,min orbital period and the magnitudes and phases
of the three significant Fourier peaks allowed them to solve for
$\Phi_{orb}$ (and therefore $x_{orb}$) and $\phi_{orb}$ directly.

Small amplitude phase modulation might also be observed from freely
precessing neutron stars \citep*[e.g.][]{nfw90, gle95b}.  For certain
geometries, the geometric wobble of the pulsar beam would periodically
modulate the pulse arrival times and cause the system to appear as a
compact binary.  If such rapidly precessing objects exist, they would
display a small number ($\sim 1-3$) of pairs of sidebands in a power
spectrum of a long observation, since the maximum possible phase
modulation amplitude for such a system is $\Phi_{orb} \lesssim
\pi/2$\,radians \citep{nfw90}.

\subsection{Elliptical Orbits} \label{sec:elliptical}

\citet{km96} derived a Fourier expansion for elliptical orbital motion
with the useful property that the magnitudes of higher order harmonics
of the orbital expansion (which correspond to independent modulation
amplitudes $\Phi_{orb,h'}$) decrease monotonically.  In the limit of
circular orbits the expansion reduces to the single phase modulating
sinusoid found in eqn.~\ref{eq:pmsig}.

Each harmonic of the orbital modulation generates sidebands around the
spin harmonic of the form discussed in \S\ref{sec:circular}.  The
sidebands created by harmonic $h'$ of the orbital expansion are
separated from each other by $h'r_{orb}$ Fourier bins.  Since
$\Phi_{orb, h'}$ monotonically decreases with increasing orbital
harmonic $h'$, the number of sidebands generated decreases while their
spacing and average amplitudes increase.  The most important aspect of
this superposition of sidebands is the fact that the Fourier response
near a spin harmonic contains sidebands spaced by $r_{orb}$ Fourier
bins --- from the fundamental of the orbital expansion --- just as in
the limiting case of circular orbits.

\section{Pulsars with Deterministic Periodic Phase Modulation} \label{sec:astro}

\begin{figure}
  \plotone{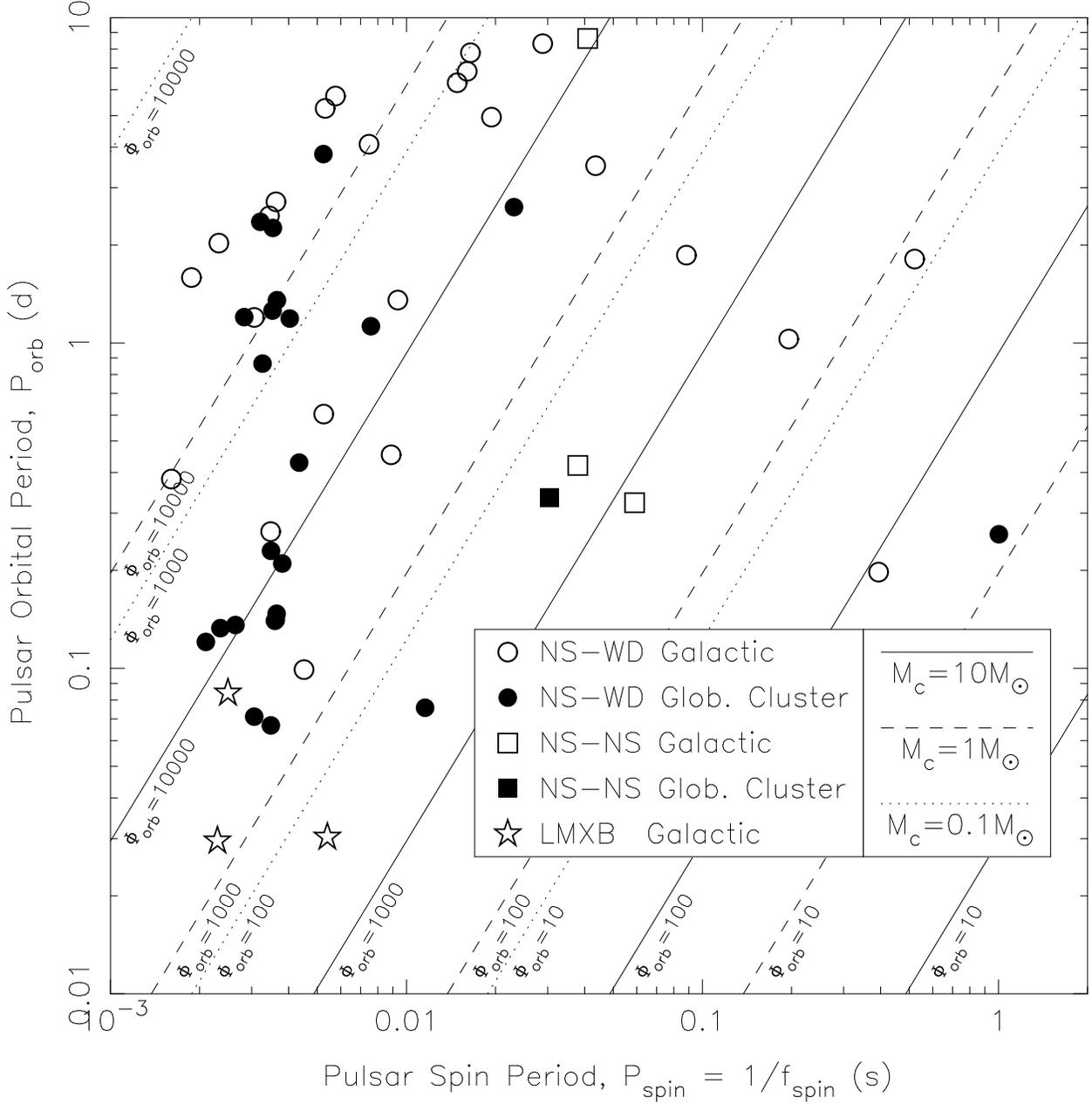}
  \caption{\footnotesize
    Orbital versus spin periods for 54 currently known binary pulsars
    with $P_{orb} < 10$\,d.  The symbols denote the type of system
    (circles for NS-WD systems, squares for NS-NS systems, and stars
    for the three recently discovered accreting X-ray MSPs) as well as
    its location in the galaxy or a globular cluster (open or filled
    respectively).  Lines of constant phase modulation amplitude
    $\Phi_{orb} = 2\pi x_{orb}f_{spin}$ are overlaid assuming orbital
    inclinations of 60\degr, pulsar masses of 1.4\,\sm, and companion
    masses of 0.1, 1, and 10\,\sm\ for dotted, dashed, and solid lines
    respectively.
    \label{fig:pvsporb}}
\end{figure}

\begin{figure}
  \plotone{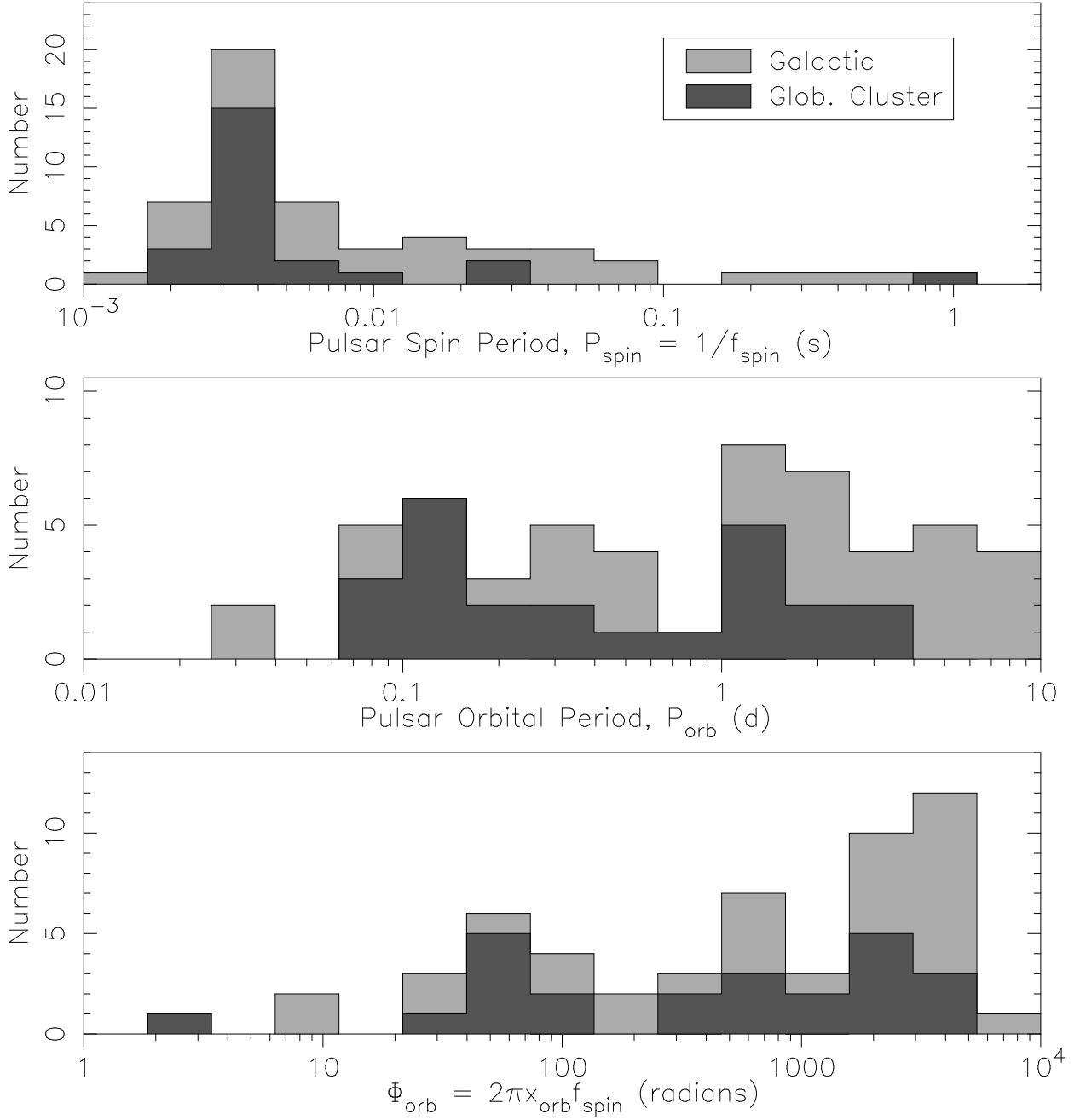}
  \caption{\footnotesize
    Histograms of spin period, orbital period, and phase modulation
    amplitude $\Phi_{orb}$ for the 54 binary pulsar systems described
    in Fig.~\ref{fig:pvsporb}.
    \label{fig:histograms}}
\end{figure}

Figures~\ref{fig:pvsporb} and~\ref{fig:histograms} show $P_{spin}$,
$P_{orb}$, and $\Phi_{orb}$ for 54 binary pulsars from the literature,
all with orbital periods less than 10 days.  A few conclusions can be
reached immediately.  First, radio millisecond pulsars (MSPs) are the
most common binary pulsar.  Second, there is a relatively flat
distribution of $P_{orb}$ for radio pulsars which cuts off quite
dramatically near $P_{orb} \sim 0.08$\,d ($\sim 2$\,h) --- the two
systems with shorter orbital periods are the recently discovered
accreting X-ray MSPs XTE~J1751-314 \citep{msszm02} and XTE~J0929-314
\citep{gcm+02}.  And third, a large fraction of the systems are
located in globular clusters.

\placefigure{fig:pvsporb}
\placefigure{fig:histograms}

The fact that large numbers of relatively compact binary pulsar
systems should contain MSPs is not surprising in light of the standard
``recycling'' model for MSP creation \citep{ver93, pk94}.  This model
explains not only the preponderance of pulsars with millisecond spin
periods, but also the fact that many MSPs have low-mass
($\sim0.02-0.4\,\sm$) companion stars in compact circular orbits.
These recycled systems make up the vast majority of the pulsars shown
in Figure~\ref{fig:pvsporb} and have typical values of $\Phi_{orb}$ of
$10^2$ to a few times $10^3$\,radians.

The cutoff in $P_{orb}$ at approximately two hours is almost certainly
due in some part to selection effects.  The rapid orbital motion of
these systems prevents their detection using conventional search
techniques --- including acceleration searches --- for all but the
brightest systems \citep{jk91}.  In fact, the only MSPs discovered
with shorter orbital periods ($P_{orb}\sim40$\,min for both
XTE~J1751-314 and XTE~J0929-314) were discovered during transient
X-ray outbursts when the pulsed fractions were very high
\citep{msszm02, gcm+02}.  Since selection effects keep us from
detecting ultra-compact systems that are not exceptionally bright, it
is difficult to determine how much the observed lack of such systems
is due to a true lack of compact systems.  It is certain that neutron
star (NS) systems with much shorter orbital periods \emph{do}\ exist,
as the new accreting X-ray MSPs and other ultra-compact LMXBs such as
4U~1820$-$303 ($P_{orb} \sim 11$\,min), 4U~1850$-$087 ($P_{orb} \sim
21$\,min), 4U~1627$-$673 ($P_{orb} \sim 42$\,min), and 4U~1916$-$053
($P_{orb} \sim 50$\,min) \citep{cgb01,lvv01} have shown.

The large fraction of known systems inhabiting globular clusters can
be explained by three basic facts.  First, due to their known location
on the sky, long targeted observations of clusters allow fainter
systems to be detected \citep[e.g.][]{and92}.  Second, repeated
observations of certain clusters have allowed many very weak pulsars
to be detected with the help of amplification from interstellar
scintillation \citep{clf+00, pdm+01}.  Finally, large numbers of
binary pulsars are expected to have been created in globular clusters
due to dynamical interactions \citep[e.g.][]{ka96, dh98, rpr00}.

Recently, \citet{rpr00} have conducted extensive population synthesis
studies for binary pulsars in dense globular clusters like 47~Tucanae.
Their initial results imply that exchange interactions among
primordial binaries will produce large numbers of NSs in binary
systems.  In fact, their results predict that large populations of
binary pulsars with low mass companions ($m_c \sim 0.05\,\sm$) should
exist in many globular clusters with orbital periods as short as $\sim
15$\,min.

The evidence seems to suggest that a large population of ultra-compact
binary pulsars should exist --- particularly in globular clusters ---
which have so far eluded detection.  With long observations, these
pulsars will be detected if search algorithms can identify the
distinctive sidebands generated by the systems.  The rest of this
paper discusses such an algorithm.

\section{Sideband Searches} \label{sec:search}

The fact that power from an orbitally modulated sinusoid is split into
multiple sidebands allows us to increase the signal-to-noise of the
detected response by incoherently summing the sidebands.  According to
one of the addition theorems of Bessel functions
\begin{equation}
  \label{eq:besseladd}
  \sum_{m=-\infty}^{\infty}J_m(z)^2 = 1,
\end{equation}
theoretically \emph{all}\ of the signal power can be recovered in this
manner for noiseless data.  Practically, though, complete recovery of
signal power is not possible for data with a finite signal-to-noise
ratio --- especially if $\Phi_{orb} \gg 1$.  Such a summation would
require calculation of the $\sim2\Phi_{orb}$ significant sidebands at
the precise Fourier frequencies of their peaks $r_{spin}\pm sr_{orb}$.
This computationally daunting task would result in very large numbers
of independent search trials when $\Phi_{orb}$, $r_{spin}$, and
$r_{orb}$ are unknown.

Even if we could recover all of the power using a sum of the
sidebands, the significance of the measurement would be less than the
significance of a non-modulated sinusoid with the same Fourier power.
This loss in significance is due to the fact that noise is co-added
along with signal.  The exact loss in significance can easily be
calculated since the probability for a sum of $m$ noise powers to
exceed some power $P_{sum}$ is the probability for a $\chi^2$
distribution of $2m$ degrees of freedom to exceed $2P_{sum}$
\citep[see e.g.][and references therein]{gro75d}.

In the case where only a few sidebands are suspected to be present
(i.e. $\Phi_{orb} \lesssim 5$), a brute-force search using incoherent
sideband summing may be computationally feasible and worthwhile.
Unfortunately, except for binary X-ray pulsars with long spin periods
and possibly certain freely precessing NSs (see \S\ref{sec:smallamp}),
the number of astrophysical systems with $\Phi_{orb} \lesssim 5$ is
probably small compared to those with $\Phi_{orb} \gg 1$ (see
\S\ref{sec:astro}).

\subsection{Two-Staged Fourier Analysis} \label{sec:twostage}

For systems with $\Phi_{orb} \gg 1$, the large number of regularly
spaced sidebands allows a very efficient detection scheme.
\emph{Since the sidebands are separated from each other by $r_{orb}$
  Fourier bins, they appear as a localized periodicity in the power
  spectrum of the original time series which can be detected with a
  second stage of Fourier analysis.}  By stepping through the full
length power spectrum and taking short Fast Fourier Transforms (FFTs)
of the powers, we incoherently sum any sidebands present and
efficiently detect their periodic nature.  As an added benefit, the
measured ``frequency'' of these ``sideband pulsations'' gives a direct
and accurate measurement of the \emph{orbital}\ period.

We define the following notation to represent the two distinct stages
of the Fourier analysis for detecting a phase modulated signal.  As
defined in eqn.~\ref{eq:pmsig}, the initial time series $n_j$ contains
$N$ points.  After Fourier transforming the $n_j$ (usually using an
FFT of length $N$), the complex responses at Fourier frequencies $r$
are represented by $A(r)$ as defined in eqn.~\ref{eq:ftdef}.  The
powers and phases are then simply $P(r)$ and $\phi(r)$ respectively.
The short FFTs of the $P(r)$ used to detect the sideband pulsations
(i.e.~the {\em secondary} power spectrum) are of length $M$ (where $M
\ll N$ and should closely match the extent of the sideband structure
in the original power spectrum) and generate complex responses
$A_2(r)$, powers $P_2(r)$, and phases $\phi_2(r)$.

In order for a pulsed signal to undergo enough modulation to produce
periodic sidebands, the observation must contain more than one
complete orbit (i.e.  $r_{orb} > 1$).  One might initially think that
two complete orbits would be necessary to create a Nyquist sampled
series of sidebands in the original power spectrum.  However, since
the sideband pulsations are not bandwidth limited, periodic signals
with ``wavelengths'' less than two Fourier bins (i.e.  $r_{orb} < 2$)
will still appear in the $P_2(r)$.  Instead of appearing at Fourier
frequency $r = M/r_{orb}$ as for signals with $r_{orb} > 2$ (see
\S\ref{sec:determination}), the fundamental harmonic of the response
will appear aliased around the Nyquist frequency ($r_{Nyq}=M/2$) at
$r_{alias} = M-M/r_{orb}$.  While $r_{orb} > 1$ is required to begin
to produce periodic sidebands, simulations have shown that this
technique realistically requires $r_{orb} \gtrsim 1.5$ to create {\em
  detectable} sidebands from most binary systems during reasonable
observations.

As the number of orbits present in the data ($r_{orb}$) increases, so
does the spacing (in Fourier bins) between the sidebands.  Since each
sideband has a traditional \sinc\ function shape in $A_2(r)$ (see
\S\ref{sec:circular}), the full-width at half-maximum (FWHM) of a
sideband is approximately one Fourier bin.  Therefore, an increase in
spacing effectively decreases the duty-cycle of the sideband
pulsations (i.e.  $\sim 1/r_{orb}$).  Signals with small duty-cycles
are easier to detect for two reasons.  First, significant higher
harmonics begin to appear in $P_2(r)$ which can be incoherently summed
in order to improve the signal-to-noise ratio.  A rule-of-thumb for
the number of higher harmonics present in $P_2(r)$ is approximately
one-half the inverse of the duty-cycle ($\sim r_{orb}/2$).  Second, as
the duty-cycle decreases, the Fourier amplitudes of each harmonic
increase in magnitude up to a maximum of twice that of a sinusoid with
an identical pulsed fraction \citep[see][for a more detailed
discussion of pulse duty-cycle effects]{rem02}.  This effect makes
each individual harmonic easier to detect as $r_{orb}$ increases.

\subsection{Search Considerations} \label{sec:recipe}

As the number of orbits present in a time series increases, the
signal-to-noise ratio of a detectable pulsar spin harmonic goes
approximately as $\sim(r_{orb}-1)^{-\beta}$ where $0.3 \lesssim \beta
\lesssim 0.4$ (see Figures~\ref{fig:sim2ms} and \ref{fig:sim20ms}).
In contrast, the signal-to-noise ratio of a pulsar spin harmonic
detected coherently goes as $T_{obs}^{-1/2}$.  This leads to a
rule-of-thumb for those conducting sideband searches: ``The longer the
observation the better.''  When making very long observations it is
important to keep the on-source time (i.e.  the window function) as
continuous and uniform as possible.  Significant gaps in the data ---
especially periodic or recurring gaps of the kind common in X-ray
observations --- create side-lobes around Fourier peaks of
constant-frequency pulsations that a sideband search will detect.  It
is a good idea before a sideband search to take the power spectrum of
the power spectrum of the known window function in order to identify
spurious orbital periods that a sideband search might uncover.

Since phase modulation sidebands require coherent pulsations (i.e.
$f_{spin}$ is constant in time) in order to form, it may be necessary
to correct the time series (or equivalently the Fourier transform)
such that the underlying pulsations are coherent.  As an example, if
the telescope's motion with respect to the solar system barycenter is
not taken into account during long observations, interstellar
pulsations --- and therefore any phase modulation sidebands as well
--- will be ``smeared'' across numerous Fourier bins due to Doppler
effects.  By stretching or compressing the time series to account for
the Earth's motion before Fourier transforming \citep[or equivalently,
by applying the appropriate Fourier domain matched filter
\emph{after}\ the Fourier transform, see][]{rem02} the smeared signal
can be made coherent for the purposes of the sideband search.

If the time-dependence of $f_{spin}$ is unknown \emph{a priori}, due
to accretion in an LMXB, the spin-down of a young pulsar, or timing
(spin) noise, for example, one could in principle attempt a series of
trial corrections and a sideband search for each trial.  Such a search
methodology is similar in computational complexity to traditional
``acceleration'' searches.

While \S\ref{sec:twostage} provided basic principles for detecting
compact binary pulsars using the two-staged Fourier analysis, in the
rest of this section we provide a more detailed discussion of how a
search might be conducted.  Figure~\ref{fig:recipe} shows two examples
of sideband searches.  The left column shows data from an 8\,h
observation of the globular cluster 47~Tucanae taken with the Parkes
radio telescope at 20\,cm on 2000 November 17.  The signal is the
fundamental harmonic of the 2.1\,ms binary pulsar 47~Tuc~J ($f_{spin}
= 476.05$\,Hz, $P_{orb} = 2.896$\,h, $r_{orb} = 2.762$, $\Phi_{orb} =
120.87$\,radians, and $m_c \sim 0.03$\,\sm) with a signal-to-noise
ratio of $\eta = a/\sigma_{n_j} \sim 0.0054$.  The right column shows
data from a simulated 8\,h observation similar to the 47~Tuc
observation described above.  The signal is the fundamental harmonic
of the 2.0\,ms binary pulsar used in Figure~\ref{fig:phsmod}\ 
($f_{spin} = 500$\,Hz, $P_{orb} = 50$\,min, $r_{orb} = 8$, $\Phi_{orb}
= 413.44$\,radians, and $m_c = 0.2$\,\sm) with a signal-to-noise ratio
one half that of the 47~Tuc~J observation ($\eta \sim 0.0027$).

\placefigure{fig:recipe}

Once a long observation has been obtained and prepared as discussed
above, we Fourier transform the data and compute the power spectrum
$P(r)$.  We then begin a sideband search which is composed of two
distinct parts: initial detection of a binary pulsar and the
determination of a detected pulsar's orbital elements.

\subsubsection{Initial Detection} \label{sec:detection}

To maximize the signal-to-noise of a detection in $P_2(r)$, we want to
match both the length and location of a short FFT $M$ with the width
and location of a pulsar's sideband response ($\sim
2r_{orb}\Phi_{orb}$) in $P(r)$.  Because a pulsar's orbital parameters
are unknown \emph{a priori}, we must search a range of short FFT
lengths $M$ and overlap consecutive FFTs.  Therefore, we must choose
the number and lengths of the short FFTs, the fraction of each FFT to
overlap with the next, and the number of harmonics to sum in each FFT.
These choices are influenced by the length of the observation and the
nature of the systems that we are attempting to find.

All currently known binary pulsars with $P_{orb}<10$\,d have
modulation amplitudes of $\Phi_{orb} \lesssim 6500$\,radians (see
Figure~\ref{fig:histograms}), with the vast majority in the range
$\Phi_{orb} \sim 20-5000$\,radians.  This range also includes many
currently unknown but predicted ``holy grail''-type systems such as
compact MSP-black hole systems or MSP-WD systems with orbital periods
of $\sim10$\,min.  This range of $\Phi_{orb}$ values is due to the
fact that the short spin periods of MSPs tend to offset the small
semi-major axes ($x_{orb}$) of their typical binary systems, while
binary systems with wider orbits tend to have longer spin periods.
Multiplying this range by twice the minimum $r_{orb}$ of 1.5 gives a
rule-of-thumb range of $M \sim 60-15000$ where the high end should be
increased based on expected values of $r_{orb}$.  For $8-12$\,h
globular cluster searches a reasonable range of values would be $64
\lesssim M \lesssim 65536$ in powers-of-two increments, for a total of
11 different values for $M$.

The choice of the overlap fraction is more subjective.  To optimize a
detection in $P_2(r)$, one should overlap the short FFTs by at least
50\% in order to avoid excessive loss of signal-to-noise when a series
of sidebands straddles consecutive FFTs.  Larger overlap percentages
may result in higher signal-to-noise, but at the expense of increased
computational costs and numbers of search trials (even though the
trials are not completely independent when overlapping).

In Figure~\ref{fig:recipe} plots $a$ and $e$, the grey shaded regions
show the sections of the $P(r)$ that were Fourier transformed for the
sideband search.  The short FFTs of length $M=1024$ and $M=8192$ were
centered on the known pulsar spin frequency and resulted in the $P_2(r)$
displayed in plots $b$ and $f$ respectively.  The short FFT lengths
were chosen as the shortest powers-of-two greater than
$2r_{orb}\Phi_{orb}$.  For 47~Tuc~J (plot $a$), the ``horned''
structure of the phase modulation sidebands described in
\S\ref{sec:circular} is easily recognizable.  This shape is not
apparent in plot $e$, and in fact, traditional searches would detect
nothing unusual about this section of the power spectrum.

Normalization of the $P_2(r)$ results in the usual exponential
distribution with mean and standard deviation of one for a transform
of pure noise.  If one first normalizes the $P(r)$ as shown in plots
$a$ and $e$ \citep[see][for several normalization techniques]{rem02},
the resulting $P_2(r)$ can be normalized simply by dividing by $M$.

Plots $b$ and $f$ of Figure~\ref{fig:recipe} show the highly
significant detections of the binary pulsars in $P_2(r)$.  For
47~Tuc~J, only one orbital harmonic is visible in $P_2(r)$ since
$r_{orb} < 2$, but the single-trial significance is approximately
$\sim 23\,\sigma$.  For the simulated 50\,min binary pulsar, 3
significant orbital harmonics are detected (predicted harmonic
locations are marked with a dotted line in plot $f$).  When summed,
the detection has a single-trial significance of $\sim 15\,\sigma$.

\subsubsection{Orbit Determination} \label{sec:determination}

Once a binary pulsar has been detected, examination of the $P(r)$ and
$\phi(r)$ allow the estimation of the Keplerian orbital parameters and
the pulsar spin period $P_{spin}$.  Brute-force matched filtering
searches centered on the estimates are then used to refine the
parameters and recover the fully coherent response of the pulsar.

The simplest parameter to determine is the orbital period $P_{orb}$.
The spacing between sidebands is $r_{orb}=T_{obs}/P_{orb}$ bins,
implying that the most significant peak in $P_2(r)$ should occur at
Fourier frequency $M/r_{orb}$.  Conversely, if we measure the Fourier
frequency $r_{meas}$ in $P_2(r)$ with the most power, we can solve for
$r_{orb}$ using $r_{orb} = M/r_{meas}$.  By using Fourier
interpolation or zero-padding to oversample $A_2(r)$ and $P_2(r)$, we
can determine $r_{meas}$ to an accuracy of
\begin{equation}
  \label{eq:rerr}
  \sigma_r = \frac{3}{\pi\alpha\sqrt{6 P_2(r_{meas})}},
\end{equation}
where $P_2(r_{meas})$ is the normalized power at the peak and $\alpha$
is the signal ``purity'' --- a property that is proportional to the
root-mean-squared (RMS) dispersion of the pulsations in time about the
centroid --- which is equal to one for pulsations present throughout
the data \citep[see][]{mdk93, rem02}.  Remembering that $P_{orb} =
T_{obs} / r_{orb}$ and solving gives
\begin{equation}
  \label{eq:porb}
  P_{orb} = \frac{T_{obs}r_{meas}}{M} \pm 
  \frac{3T_{obs}}{\pi\alpha M\sqrt{6P_2(r_{meas})}}.
\end{equation}
For aliased signals, the orbital period is
\begin{equation}
  \label{eq:porbalias}
  P_{orb, aliased} = \frac{T_{obs}\left(M-r_{meas}\right)}{M} \pm 
  \frac{3T_{obs}}{\pi\alpha M\sqrt{6 P_2(r_{meas})}}.
\end{equation}
For pulsars with $\Phi_{orb} \gg 1$ observed such that $r_{orb} > 2$,
it is often possible to measure $P_{orb}$ to one part in $10^{4}$,
corresponding to a $\sim 1$\,s error for many of the pulsars shown in
Figure~\ref{fig:histograms}.

\begin{figure}
  \epsscale{0.8}
  \plotone{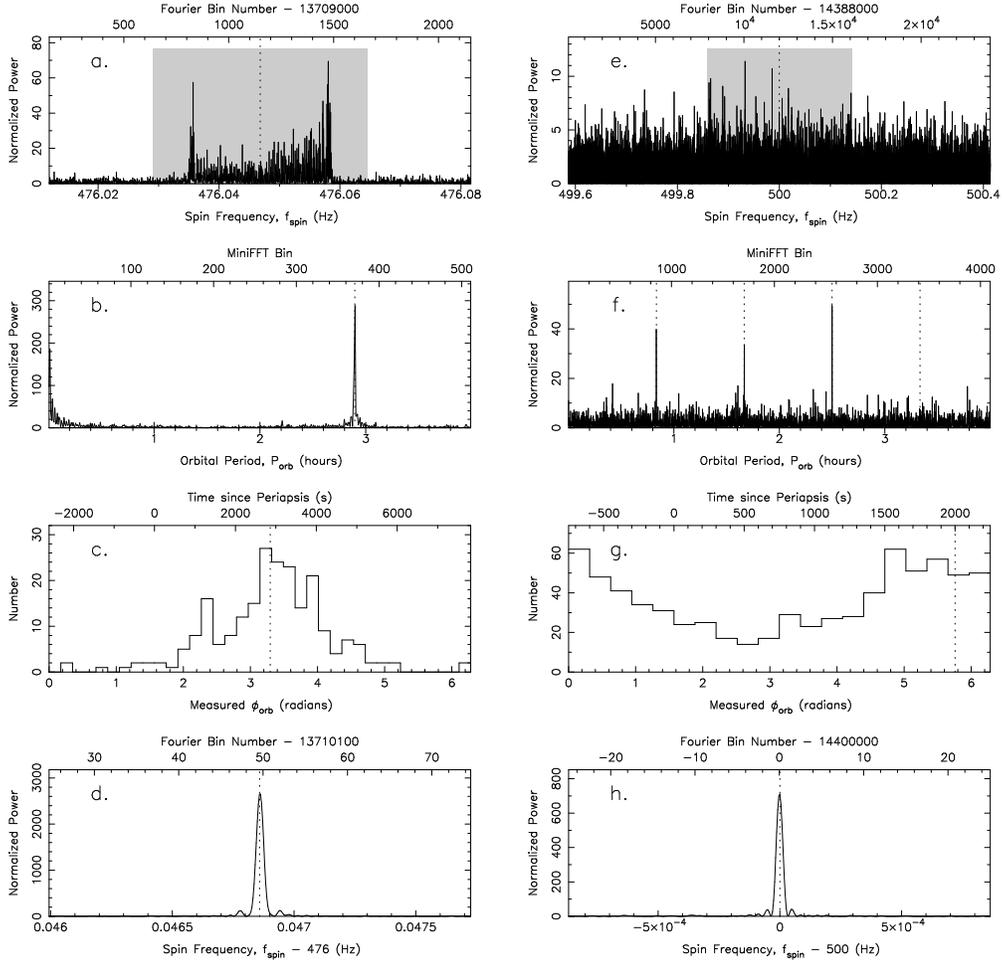}
  \caption{\footnotesize
    Two examples of key parts of a phase modulation sideband search.
    The left column shows the analysis of an 8~h observation of the
    globular cluster 47~Tucanae taken with the Parkes radio telescope
    of the binary millisecond pulsar 47TucJ.  The right column shows
    analysis of simulated data containing a weaker millisecond pulsar.
    Plots $a$ and $e$ show portions of the full length power spectra
    $P(r)$ centered on the fundamental spin harmonics $r_{spin} =
    f_{spin}T_{obs}$ of the pulsars.  The grey regions are the
    sections of the power spectra $P(r)$ that were Fourier analyzed in
    order to create the power spectra $P_2(r)$ shown in plots $b$ and
    $f$.  Plots $c$ and $g$ show histograms of the orbital phase as
    measured using the raw Fourier amplitudes in the original FFTs as
    described in \S\ref{sec:determination}.  Once the three Keplerian
    elements for circular orbits were determined, a Fourier domain
    orbital template was calculated and correlated with the raw
    Fourier amplitudes (plots $d$ and $h$).  This process recovered
    all of the power from the pulsar's fundamental spin harmonic that
    had initially (plots $a$ and $e$) been spread over many Fourier
    bins.  The dotted lines in each plot show the known values of the
    independent variable.  A more detailed description of each plot is
    given in \S\ref{sec:recipe}.
    \label{fig:recipe}}
  \epsscale{1.0}
\end{figure}

Once $P_{orb}$ is known it becomes possible to measure the projected
semi-major axis $x_{orb}$ by determining the total width of the
sidebands in $P(r)$.  For bright pulsars like 47~Tuc~J, the number of
Fourier bins between the ``horns'' ($\Delta_r \sim
2\Phi_{orb}r_{orb}$) as shown in Figure~\ref{fig:recipe}$a$ can be
measured directly and converted into $x_{orb}$ using
\begin{equation}
  \label{eq:xorb}
  x_{orb} \simeq \frac{\Delta_r}{4\pi f_{spin}r_{orb}}, 
\end{equation}
where $f_{spin}$ can be estimated as the frequency midway between the
phase modulation ``horns''.

For weaker pulsars, where the sideband edges are not obvious (as in
Figure~\ref{fig:recipe}$e$), $\Delta_r$ can be estimated using
numerous short FFTs.  By taking a series of short FFTs of various
lengths around the detection region of $P(r)$ and measuring the power,
``centroid'' and ``purity''\footnote{The measured values of
  ``centroid'' and ``purity'' are estimates of a signal's location and
  duration in a time series as determined from the derivatives of the
  Fourier phase and power at the peak of the signal's response
  \citep[see][]{mdk93, rem02}.} at the frequency corresponding to
$P_{orb}$, one can map the extent of the hidden sidebands and compute
an estimate for $x_{orb}$.

For pulsars in circular orbits, which constitute the majority of the
systems in Figures~\ref{fig:pvsporb} and \ref{fig:histograms}, the
only remaining orbital parameter is the time of periapsis passage
$T_o$.  Defining the time since periapsis passage as $\Delta_{T_o} =
t_o - T_o$ and using eqn.~\ref{eq:phiorb} we get
\begin{equation}
  \label{eq:deltaTo}
  \Delta_{T_o} = \frac{P_{orb}}{2\pi}\left(\phi_{orb}-\frac{\pi}{2}\right), 
\end{equation}
where $\phi_{orb}$ can be measured from the phases of the sidebands
in $A(r)$ using the following technique.  

From eqn.~\ref{eq:pmsimple} we know that the sidebands have phases of
$\phi_s = s\left(\phi_{orb}+
  \frac{\pi}{2}\right)+\phi_{spin}$\,radians.  Unfortunately, some of
the {\em measured} phases will differ by $\pi$\,radians from those
predicted by eqn.~\ref{eq:pmsimple} since the corresponding sideband
amplitudes (the $J_s(\Phi_{orb})$) are negative (see footnote 5).  If
we {\em predict} values for $J_s(\Phi_{orb})$ using our knowledge of
$r_{orb}$, $f_{spin}$, and $x_{orb}$, we can ``flip'' (i.e.~ add or
subtract $\pi$\,radians to) the measured phases of the sidebands that
are predicted to have negative amplitudes.  We can then estimate
$\phi_{orb}$ simply by subtracting the phases of neighboring sidebands
using $\phi_{orb} = \phi_{s+1} - \phi_s - \pi/2$.

For weak pulsars each measurement of $\phi_{orb}$ has a large
uncertainty.  Fortunately, we can make $\sim2\Phi_{orb}$ measurements
of $\phi_{orb}$ by using each pair of sidebands and then determine
$\phi_{orb}$ and its uncertainty statistically.
Figure~\ref{fig:recipe} plots $c$ and $g$ show histograms of the
measurement of $\phi_{orb}$ using this technique.

It is important to mention that for weak pulsars it may be difficult
to determine the location of the sidebands in order to measure their
phases.  The location of the sideband peaks can be calculated by
measuring the phase $\phi_{meas}$ of the fundamental harmonic
discovered in $A_2(r)$.  The number of Fourier bins from the first bin
used for the short FFT to the peak of the first sideband in $A(r)$ is
simply $\phi_{meas}r_{orb}/2\pi$.  Subsequent peaks are located at
intervals of $r_{orb}$ Fourier bins.

Once estimates have been made for the Keplerian orbital parameters, we
compute a set of complex sideband templates (i.e.~matched filters for
the Fourier domain response of an orbitally modulated sinusoid) over
the most likely ranges of the parameter values given the uncertainties
in each.  These templates are then correlated with a small region of
the Fourier amplitudes $A(r)$ around $f_{spin}$.  When a template
matches the pulsar sidebands buried in $A(r)$, we recover essentially
all the power in that pulsar spin harmonic \citep{rem02}.
Figure~\ref{fig:recipe} plots $d$ and $h$ show the results of just
such a matched filtering operation.  The Doppler effects from the
orbital motion have been completely removed from the data and the
resulting Fourier response is that of an isolated pulsar.

For binary pulsars in eccentric orbits, the techniques discussed here
could in principle be applied to each set of sidebands from the
orbital Fourier expansion (see \S\ref{sec:elliptical}).  Such an
analysis would be much more complicated than that described here and
would require a high signal-to-noise detection.

\section{Discussion} \label{sec:discussion}

We have described a new search technique for binary pulsars that can
identify sidebands created by orbital modulation of a pulsar signal
when $T_{obs} > P_{orb}$.  Sideband or phase-modulation searches allow
the detection of short orbital period binary pulsars that would be
undetectable using conventional search techniques.

\begin{figure}
  \plotone{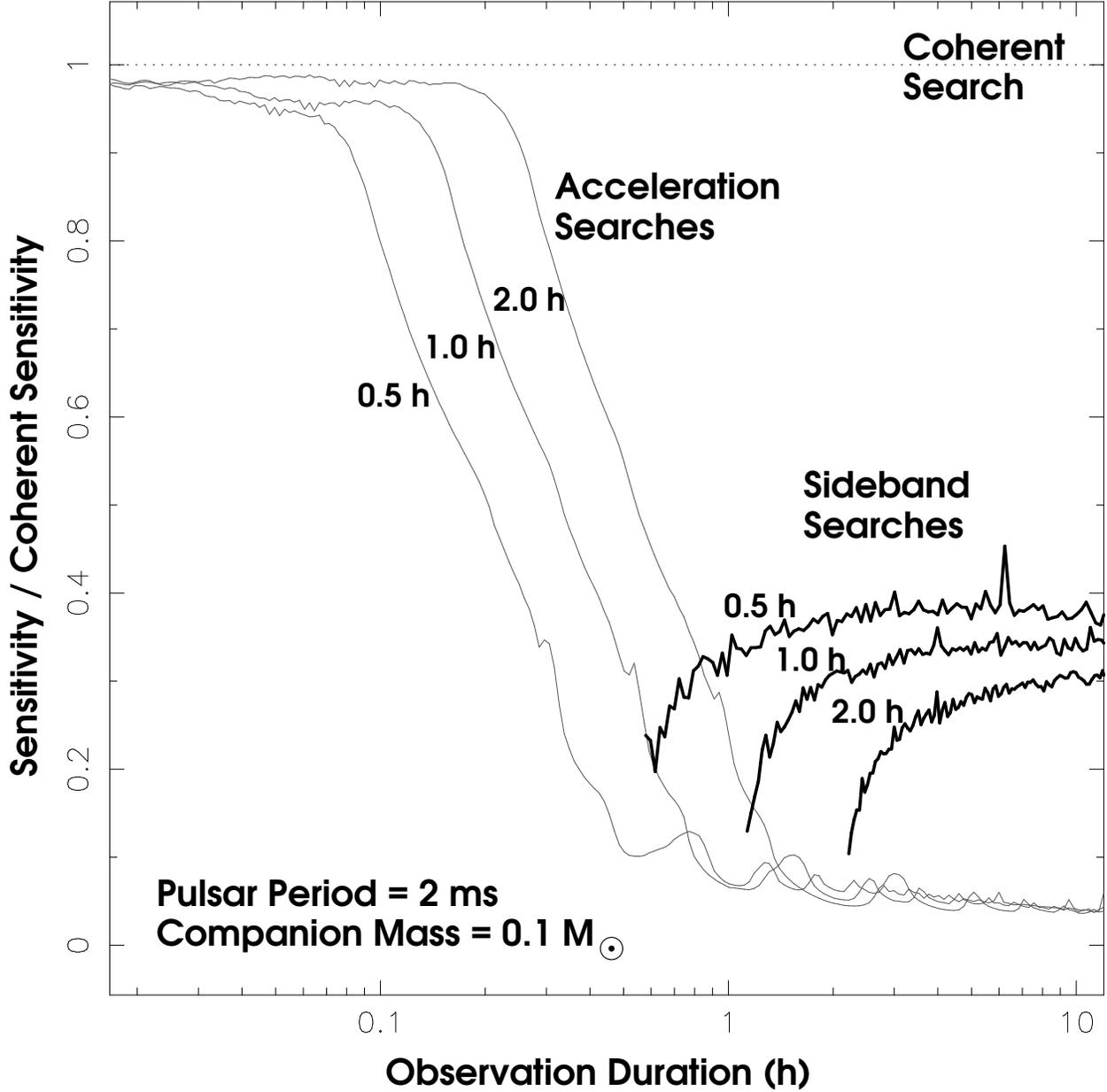}
  \caption{\footnotesize
    The results of simulations showing the sensitivity of both
    acceleration and sideband searches for the fundamental harmonic
    (i.e. a sinusoid) of a 2\,ms binary pulsar with a 0.1\,\sm\ 
    companion and orbital periods of 0.5, 1, and 2\,h as a fraction of
    the optimal (i.e. coherent) sensitivity, which is proportional to
    $T_{obs}^{-1/2}$.  The detection threshold used in the simulations
    was $8\,\sigma$.  Sideband searches are significantly more
    sensitive to binary pulsars than optimal duration acceleration
    searches when $T_{obs} \gtrsim 2P_{orb}$.
    \label{fig:sim2ms}}
\end{figure}

\begin{figure}
  \plotone{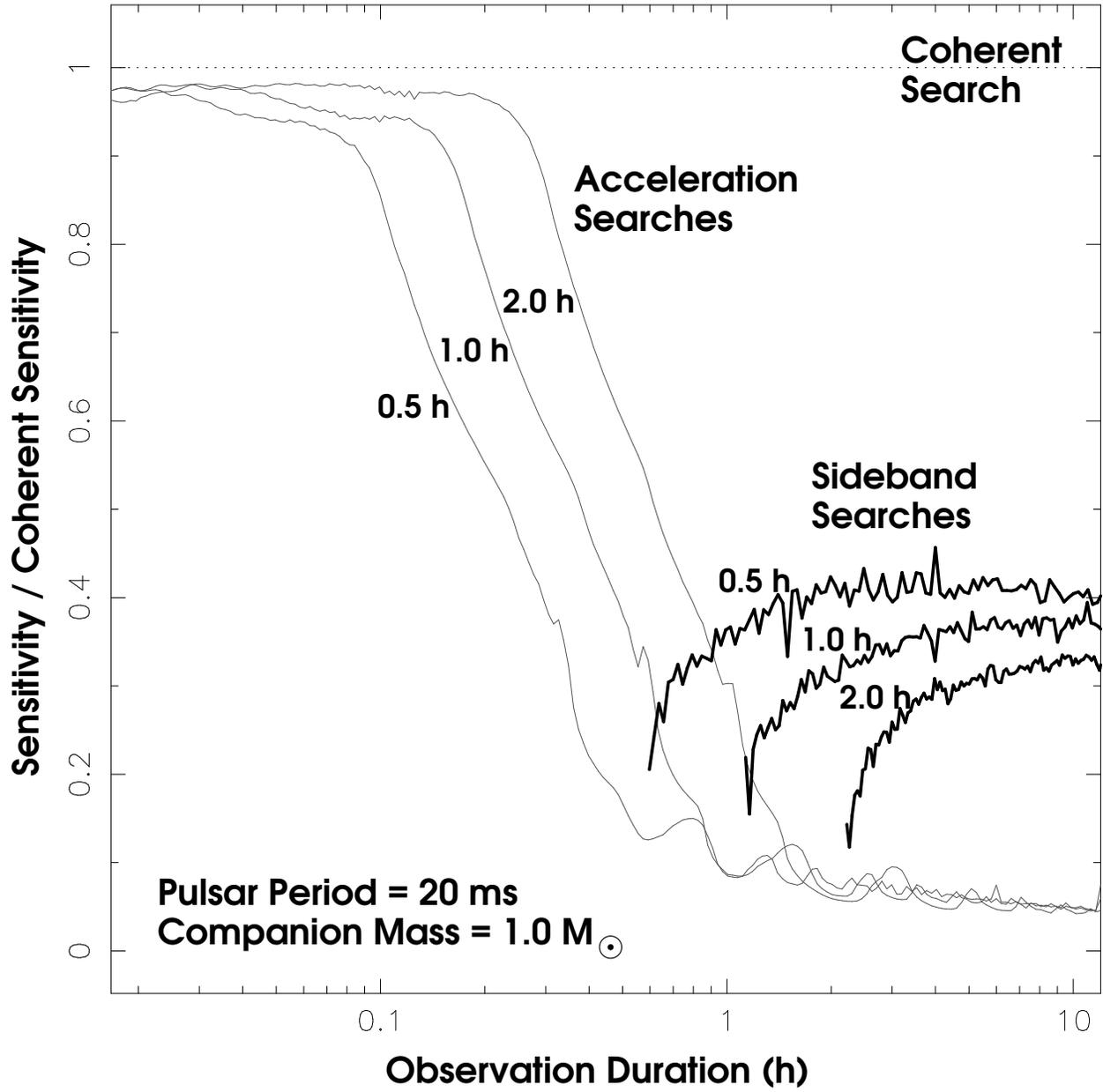}
  \caption{\footnotesize
    Results from simulations similar to those shown in
    Figure~\ref{fig:sim2ms} but for a 20\,ms binary pulsar with a
    $1.0$\,\sm\ companion.
    \label{fig:sim20ms}}
\end{figure}

Figures~\ref{fig:sim2ms} and \ref{fig:sim20ms} show how the
sensitivity of sideband searches compares to that of acceleration
searches for a wide range of observation times $T_{obs}$.
Acceleration search sensitivities are near-optimal when $T_{obs}
\lesssim 0.1 P_{orb}$ \citep{jk91}, but degrade rapidly at longer
integration times when the constant frequency derivative approximation
breaks down.  Sideband searches, on the other hand, require $T_{obs}
\gtrsim 1.5 P_{orb}$ in order to approach the sensitivity of optimal
duration acceleration searches, but for longer $T_{obs}$ sensitivity
improves as $\sim(r_{orb}-1)^{-\beta}$ where $0.3 \lesssim \beta
\lesssim 0.4$.  \emph{For targeted searches of duration
  $\sim$8$-$12\,h (e.g. typical globular cluster observations),
  sideband searches are $\sim$2$-$10 times more sensitive to compact
  binary pulsars ($P_{orb} \lesssim 4$\,h) than optimal duration
  acceleration searches.}  In general, one should think of
acceleration and sideband searches as complementary to each other ---
sideband searches allow the detection of ultra-compact binary pulsars,
while acceleration searches maximize the detectability of isolated and
longer-period binary pulsars.

\placefigure{fig:sim2ms}
\placefigure{fig:sim20ms}

The fact that sideband searches target a different portion of orbital
parameter space than acceleration searches and yet require
significantly less computation time, provides good reason to include
them in future pulsar searches where $T_{obs} \gtrsim 30$\,min.  In
fact, the Parkes Multibeam Pulsar Survey \citep[$T_{obs} \sim
35$\,min,][]{lcm+00,mlc+01} has included a sideband search in an
on-going re-analysis of their survey data at the cost of only a
marginal increase in computer time.  While the probability of
detecting a binary pulsar with $P_{orb} \lesssim 23$\,min is almost
certainly quite low, the discovery of a single such system would
provide a wealth of scientific opportunity.

\acknowledgments
 
\emph{Acknowledgments}\ \ We would like to thank G.~Fazio for
supporting our research and encouraging us in this work.  Many of the
computations for this paper were performed on equipment purchased with
NSF grant PHY 9507695.  SMR~is a Tomlinson Fellow, JMC is supported by
NSF Grant 9819931 and by the National Astronomy and Ionosphere Center,
and SSE is supported by a NSF CAREER Grant.
 

\clearpage

\end{document}